\documentstyle[11pt]{article}

\begin{document}
\newcommand{\ECM}{\em Departament d'Estructura i Constituents de la
Mat\`eria \\
and\\
I. F. A. E.\\
Facultat de F\'\i sica, Universitat de Barcelona \\
Diagonal 647, E-08028 Barcelona, Spain}

\def\thefootnote{\fnsymbol{footnote}}
\pagestyle{empty}
{\hfill \parbox{6cm}{\begin{center} quant-ph/9904xxx\\
                                    April 1999\\
                     \end{center}}}
\vspace{1.5cm}

\centerline{
\large{\bf Optimal generalized quantum measurements}}
\centerline{
\large{\bf for
arbitrary spin systems}}

\vskip .6truein
\centerline {
A. Ac\'\i n, 
J.I. Latorre,
P. Pascual}

\vspace{.3cm}
\begin{center}
\ECM
\\
\end{center}
\vspace{1.5cm}

\centerline{\bf Abstract}
\medskip

Positive operator valued measurements
on a finite number of $N$ identically prepared systems
of arbitrary spin $J$ are discussed. Pure states 
are characterized in terms of  Bloch-like vectors restricted by a
 $SU(2 J+1)$ covariant constraint. 
This representation allows for a simple description of the equations
to be fulfilled by optimal measurements. We explicitly find
the minimal POVM for the  $N=2$ case,  a rigorous bound for $N=3$
and set up the analysis for arbitrary $N$.

\newpage                                                                
\pagestyle{plain}
\section{Introduction}

A measurement on a quantum mechanical system only provides partial
information on the measured state. Even in the case where $N$ identical
copies of the system are available, the information which can be
retrieved remains bounded. This fact can be quantified using the 
averaged fidelity based on the following general idea. Given $N$
identical copies of a system we may consider a two-step procedure to 
rate the fidelity of a measuring
apparatus. First, we set up a generalized quantum mechanical
measurement (or positive operator valued measurement, POVM \cite{N,P}).
 Upon
performing a measurement,  
its outcome
provides the basis for  a best guess about the incoming state. 
The averaged fidelity quantifies how
close the final guess is from the original state averaging over
the latter. For any finite number
 $N$ of copies of a spin $J$ 
pure state system the average fidelity is proven to be 
bounded by \cite{BM}
\begin{equation}                                                                     
\overline {f}(N,J)={{N+1}\over {N+2 J+1}}.
\label{fmax}
\end{equation}
 The issue at stake remains to device the  optimal and
minimal measuring strategy  for any quantum system.

Explicit constructions of optimal and minimal
generalized quantum mechanical measurements of
spin ${1\over 2}$ systems have been presented recently in Refs. 
\cite{MP,H,DBE,LPT,VLPT}.  The detailed
construction is subtle and depends  
on  whether  the original system is in a pure or mixed state.
 The simplest case corresponds to 
measuring a spin ${1\over 2}$
system known to be in a pure state. A generalized measurement
can be constructed as a resolution of the identity made with 
rank one hermitian operators, which are in turn built from
the direct product of a given state, 
\begin{equation}                                                                     
I= \sum_{r=1}^n c^2_r | \Psi_r\rangle^N\ {}^N\langle\Psi_r|\ ,
\label{identity}
\end{equation}
where $I$ is then the identity in the maximal spin subspace.
The important --and of possible future practical relevance-- result
is that the maximum averaged fidelity is attained with a
finite number of operators \cite{DBE}. Upon a case by case 
 analysis, it is found
that  the minimum number, $n$,  of such 
operators is  a function of $N$
\begin{table}[h]
\begin{center}
\begin{tabular}{|c|c|c|c|c|c|}
\hline
$N$&1&2&3&4&5\\
\hline
$n$&2&4&6&10&12\\ 
\hline
\end{tabular}
\end{center}
\end{table}
\noindent
 and is given in the table.
The explicit form of Eq. (\ref{identity}) for the above
cases can be found in Ref. \cite{LPT}. 

The far more involved case of spin ${1\over 2}$ mixed states
has also been worked out in Ref. \cite{VLPT}. At variance with the
pure state case,
the closed  expression 
for the maximum averaged fidelity depends on what the 
unbiased a priori distribution of density matrices is. Yet, explicit
solutions for optimal measurements are found. Some remarkable properties
emerge along the new construction. Let us briefly mention a few. Optimal 
measurements turn out to be structured using projectors on 
total spin eigenspaces and, within each  eigenspace, on maximal spin
component is some direction. This allows for a reuse of minimal and
optimal results from the pure state case. Also, beyond two copies,
some projectors are not  of rank one. 

Explicit constructions of optimal minimal measurements are so far
restricted to spin ${1\over 2}$ systems, either pure or mixed. It is
the purpose of this paper to extend this analysis for 
arbitrary spin pure states. A number of non-trivial issues must
be faced at the outset. For instance, progress in the spin ${1\over 2}$
case was triggered by the appropriate use of the Bloch vector 
labelling of density matrices associated to spinors.
We shall resort to a similar representation in the case of arbitrary
spin states, using representations of $SU(2 J+1)$. The equivalent of
a Bloch vector will be shown to obey a covariant restriction. 
This extra work will allow for a unified general setting of the
problem of optimal measurements of arbitrary spins. 

Finding explicit minimal optimal measurements remains a matter 
of case by case analysis. We shall provide explicit bounds for
the minimal number of projectors, $n$, in POVMs. The case of
$N=2$ will be fairly complete. Higher number of copies still need
further ingenuity  to get rigorous bounds.
                        
\section{ Averaged fidelity}

Consider a spin $J$ particle which is in an unknown 
 pure state $|\Psi\rangle$, 
\begin{equation}
  |\Psi\rangle = \left(\matrix{
      x_1+{\rm i} y_1\cr
      x_2+{\rm i} y_2\cr
      \dots\cr
      x_D+{\rm i} y_D\cr
  }\right)\ ,
\label{spinor}
\end{equation}
where $D=2 J+1$ and the  normalization of the state imposes
$\sum_{i=1,..,D}\left(x_i^2+y_i^2\right)=1$. Of course, 
we may use a different parametrization, {\sl e.g.}
\begin{equation}
  |\Psi\rangle = \left(\matrix{
    \cos\phi\cr
    \sin\phi \left(x_2+{\rm i} y_2\right)\cr
    \dots\cr
    \sin\phi \left(x_D+{\rm i} y_D\right)\cr
  }\right)\ ,
\label{spinordos}
\end{equation}
with $0\leq \phi\leq {\pi\over 2}$ and 
$\sum_{i=2,..,D}\left(x_i^2+y_i^2\right)=1$. Using this second 
parametrization and 
following Ref. \cite{SAC} 
it is  possible to prove that 
the volume element in the space of these states is
\begin{equation}
  dV_D = 4\left( \sin \phi\right)^{2D-3}\ \cos\phi \ d\phi\ dS_{2D-3}\ ,
\label{volumeleement}
\end{equation}
where $dS_{2D-3}$ corresponds to the standard volume element 
on $S_{2 D-3}$. The total volume is
\begin{equation}
  V_D = {4\pi^{D-1}\over (D-1)!}\ .
\label{volume}
\end{equation}

Given $N$ identical copies of the arbitrary spin state we
have
\begin{equation}
\vert\Psi\rangle^N\equiv \vert\Psi\rangle
\otimes\vert\Psi\rangle
\otimes...{}^N...\otimes\vert\Psi\rangle.
\label{psiN}
\end{equation}
A measurement on this enlarged system will bring richer information
on $\vert\Psi\rangle$ than $N$ separate measures on its respective copies
\cite{PW}.

Setting a generalized quantum measurements  consists in providing a
resolution of the identity of the type
\begin{equation}
\sum^n_{i=1}\ c^2_r\ \vert\Psi_r
\rangle^{N}{}^{N}\langle\Psi_r\vert+ P_N= I\ ,
\label{povm}
\end{equation}
where $P_N$ is the projector on the space different
from the one spanned from states of the form given in Eq. (\ref{psiN}).
We already have all the necessary elements to define and compute
the averaged fidelity.  Upon measuring $\vert\Psi\rangle^N$ with
the above POVM, a given outcome labelled by $r$ will result with
probability $\vert ^N\langle\Psi\vert\Psi_r\rangle^N\vert^2$. The
natural guess for the initial pure state is, then, $\vert
\Psi_r\rangle$
(this is only the best strategy if
the initial state is known to be pure;
the best guess for a mixed state is not the same state as the outcome
of the POVM \cite{VLPT}).  
 The overlap of this guess with the original
state is just $\vert \langle\Psi\vert\Psi_r\rangle\vert^2$. The
averaged or mean fidelity is defined as the product of the probability for
$r$ being triggered
 times the overlap between the ensuing guess and 
the original state, averaged over 
all possible initial unknown states,
\begin{equation}
  \overline f(N,J)\equiv {1\over V_{2 J+1}} \sum_{r=1}^n c_r^2
  \int_0^{\pi\over 2} d\phi\ \left(\sin\phi\right)^{4 J-1}\cos\phi
  \int dS_{4J-1} \left| ^N\langle\Psi\vert\Psi_r\rangle^N\right|^2
  \ \left| \langle\Psi\vert\Psi_r\rangle\right|^2\ .
\label{averagefid}
\end{equation}
To evaluate the above expression  it is convenient to use the 
freedom to choose the integration variables to set each individual
$\vert \Psi_r\rangle$ as a spinor with only nonvanishing 
first component. Then,
\begin{equation}
  \overline f(N,J)= {1\over V_{2 J+1}} \sum_{r=1}^n c_r^2
  \int_0^{\pi\over 2} d\phi\ \left(\sin\phi\right)^{4 J-1}
  \left(\cos\phi\right)^{2 N+3} S_{4 J-1}\ .
\label{paso}
\end{equation}
We finally get
\begin{equation}
  \overline f(N,J)= {(2 J)! (N+1)!\over (2 J+N+1)!}\ 
  \sum_{r=1}^n c_r^2\ .
\label{pasodos}
\end{equation}
This sum is easily calculated. It is just the dimension of the
space spanned by the totally symmetric tensor of order $N$
whose indices can take $2J+1$ values,
\begin{equation}
 \sum_{r=1}^n c_r^2 = {(2 J+N)!\over N! (2J)!}\ .
\label{sumacr}
\end{equation}
Thus,
\begin{equation}
  \overline f(N,J)= {N+1\over N+2J+1}\ ,
\label{averagefidagain}
\end{equation}
which corresponds to Eq. (\ref{fmax}) and was obtained in
Ref. \cite{BM} using different techniques.

\section{Generalized Bloch form  of arbitrary spin pure states}

 It is sometimes uselful
to represent the state of a spin ${1\over 2}$
system using the  Bloch representation, 
\begin{equation}
  \rho = {1\over 2} I+ {1\over 2} \vec b \cdot \vec \sigma \ ,
\label{blochform}
\end{equation}
where $\vec b$ is a  vector living within the unit sphere. Pure 
states correspond to the surface of the sphere, that is $\vec b^2=1$.
A similar but more complicated construction is possible for arbitrary
spin particles. 

Consider a pure state of a spin $J$ particle. One may represent it using
{\sl e.g.} Eq. (\ref{spinor}). Alternatively we may construct its
associated density matrix and write 
\begin{equation}
  \rho= {1\over 2 J+1}I + \sqrt{J\over 2 J+1}\   n_a \lambda_a \qquad 
  a=1,\dots,4J(J+1) \ ,
\label{genbloch}
\end{equation} 
where $\lambda_a$  are the 
generators of the $SU(2J+1)$ normalized by
\begin{equation}
  {\rm Tr}\left(\lambda_a \lambda_b\right) = 2 \delta_{ab}\ ,
\label{normlambdas}
\end{equation}  
and $\hat n$ is the normalized vector that plays the role of a 
generalized Bloch vector. The coefficients
in Eq. (\ref{genbloch})  are chosen in such a way that 
$  {\rm Tr} \rho={\rm Tr} \rho^2 = 1$.

A simple counting of degrees of freedom shows that a spin $J$ pure
state is described by $4J$ real parameters whereas the generalized Bloch
vector carries $4J(J+1)-1$. A mismatch appears for $J>{1\over 2}$
which implies that severe
constraints must limit the subspace of valid vectors $\hat n$. 
Indeed, pure states must verify $\rho=\rho^2$ which translates into
\begin{equation}
d_{abc} n_a n_b = {2J-1\over\sqrt{ J(2J+1)}} n_c \ ,
\label{compact}
\end{equation}
when Eq. (\ref{genbloch}) is used and
where $d_{abc}$ are the completely symmetric symbols associated to 
$SU(2 J+1)$, defined through the anticommutator fo the generators of
the group \cite{PT}
\begin{equation}
  \{ \lambda_a,\lambda_b\} = {4\over 2J+1} \delta_{ab} I + 2 \ d_{abc}
  \lambda_c 
\label{defd}
\end{equation}
which verify 
\begin{equation}
   d_{abb}=0\quad,\quad d_{abc} d_{dbc} = {(2J-1)(2J+3)\over 2J+1}
  \delta_{ad}\ .
\label{dprop}
\end{equation}

Some useful properties of the vectors $\hat n$ follow from 
the above general covariant constraint (\ref{compact}),
\begin{eqnarray}
  && d_{abc}  n_a  n_b  n_c = {2 J-1\over \sqrt{J(2J+1)}}\ ,
  \nonumber\\
  && d_{abe} d_{cde}  n_a
  n_b n_c n_d = { (2 J-1)^2\over J(2J+1)} \ ,
\label{dnnn}
\end{eqnarray}
where it is clear that for spin $J={1\over 2}$ the simple
structure of $SU(2)$ makes the $d$-symbols to vanish and the
r.h.s. to be identically zero.

We can also
deduce the useful constraint which follows from the positivity of 
the square of the scalar product of two arbitrary spin $J$ pure states
which reads
\begin{equation}
  \vert\langle \Psi\vert\Psi'\rangle\vert^2 = {\rm Tr}\left(\rho
  \rho'\right)= {1\over 2J+1} \left(1+ 2 J \hat n\cdot\hat n'\right)
  \geq 0\ .
\label{scalarproduct}
\end{equation}
Generalized Bloch vectors are thus constrained to have scalars products
bounded by
\begin{equation}
  \hat n\cdot \hat n' \geq -{1\over 2J}\ .
\label{boundscalarproduct}
\end{equation}
Two pure states are orthogonal then when the scalar product of their
generalized Bloch vectors satisfies the equality in Eq.
(\ref{boundscalarproduct}).

Let us illustrate the construction of a Bloch vector for 
the $J=1$ example. In this case, 
the  density matrix representing the system can be connected to the 
standard spinor-like representation. For instance, taking $J=1$ it is
easy to see that the generalized Bloch vector corresponds to 
Eq. (\ref{spinor}) if
\begin{eqnarray}
  n_1&=&\sqrt{3} \left(x_1 x_2+y_1y_2\right) \qquad
  n_2=\sqrt{3} \left(x_1 y_2-x_2y_1\right)\nonumber \\
  n_4&=&\sqrt{3} \left(x_1 x_3+y_1y_3\right) \qquad
  n_5=\sqrt{3} \left(x_1 y_3-x_3y_1\right)\nonumber \\
  n_6&=&\sqrt{3} \left(x_2 x_3+y_2y_3\right) \qquad
  n_7=\sqrt{3} \left(x_2 y_3-x_3y_2\right)\nonumber \\
  n_3&=&{\sqrt{3}\over 2} \left(x_1^2+y_1^2-(x_2^2+y_2^2)\right) \qquad
  n_8={1\over 2} \left(1- 3(x_3^2+y_3^2)\right)
\label{explicitbloch}
\end{eqnarray}
and $\lambda_a$ are taken in the Gell-Mann representation of 
$SU(3)$ \cite{PT}. Note that symmetric and antisymmetric combinations
of the spinor components build the raising and lowering generators,
whereas the Casimir combinations correspond to diagonal ones.
Generalization of this construction for arbitrary spin $J$ based
on the $SU(2J+1)$ group is straigthforward.

The advantage of using a generalized Bloch representation
for arbitrary spin pure states will become apparent shortly, when
all our equations will be manifestly $SU(2J+1)$ covariant and real.
This is equivalent to note that the
difference between working with spinors, which live in the fundamental
representation of the group, or with Bloch vectors, which live
in the adjoint representation, is that the second is real.

\section{Optimal measurements for a single copy of a system}

Let us go back to the construction of a generalized quantum measurement
of arbitrary spin systems. We basically need to solve for the minimal set 
of $\vert\Psi_r\rangle$
states such that Eq. (\ref{povm}) is fulfilled. We have found 
convenient to project out the $P_N$ piece using
\begin{equation}
  \sum^n_{r=1} \ c^2_r\  \left| ^N\langle\Psi\vert\Psi_r\rangle^N\right|^2
  =1\qquad \forall \vert\Psi\rangle\ \ ,
\label{basiceq}
\end{equation}
This equation  can be also written in the Bloch
representation as
\begin{equation}
  \sum^n_{r=1} \ c^2_r {1\over (2J+1)^N} \left(1 + 2J\sum_{a}\ n_a n_a(r)
  \right)^N = 1\ ,
\label{blocheq}
\end{equation}
where every $\hat n(r)$  corresponds to a pure state in the POVM and
$\hat n$ to the original pure state.

It is clear that the simplest situation we may face
corresponds to having a single copy of the unknown state. The optimal
and minimal measurement for such a case is, of course, known to
correspond to a von Neumann measurement. We shall, though, proceed
in a more general way and set the
{\sl modus operandi} for the more elaborate cases as deviced in
Ref. \cite{LPT}.

The equation (\ref{basiceq}) with $N = 1$ can be demonstrated
(with a little effort) to be
equivalent to
\begin{eqnarray}
  &&\sum^n_{r=1} \ c^2_r\ \left( x_j(r)
   x_k(r)+y_j(r)y_k(r)\right)=\delta_{jk}\nonumber\\
  &&\sum^n_{r=1} \ c^2_r\ \left( x_j(r)
   y_k(r)-x_k(r)y_j(r)\right)=0 \qquad j,k=1,\dots,2J+1
\label{previouseq}
\end{eqnarray}
Using the insight given by Eq. (\ref{blocheq}) and the result of
Eq. (\ref{sumacr}),
 this set of $(2J+1)^2$  independent 
equations can be rewritten in terms of the  Bloch vector
as
\begin{eqnarray}
  &&  \sum^n_{r=1} \ c^2_r\ = 2J+1 \nonumber\\
  &&  \sum^n_{r=1} \ c^2_r\  n_a(r) =0\ ,
\label{firstset}
\end{eqnarray}
where it is important to remember the constraints 
limiting $\hat n(r)$. For instance, scalar products between 
any pair $\hat n(r)\cdot\hat n(s) \geq -{1\over 2J}$, thus
\begin{equation}
   \sum_{r\not= s} c_r^2 \left({1\over 2J}+ \hat n(r)\cdot\hat n(s)
   \right) \geq 0\ .
\label{basicnone}
\end{equation}
Using the set of equations (\ref{firstset}) the above inequality
can be transformed into
\begin{equation}
  1- c_s^2\geq 0\qquad \forall s=1,\dots,n
\label{cscosntraint}
\end{equation}
Suming over all $s$ we get 
\begin{equation}
  n\geq 2 J+1
\label{none}
\end{equation}
This bound is indeed saturated by a von Neumann measurement,
that is 
\begin{eqnarray}
  n_{min}&=&2 J+1\nonumber\\
  c_s^2=1\ \ \forall s\quad \quad &,&\quad 
  \hat n(r)\cdot\hat n(s)=-{1\over
  2J}\ \ \forall r\not= s
\label{nonesol}
\end{eqnarray}
The explicit standard construction for $J=1$  is
recovered as the solution to this $N=1$ POVM
\begin{equation}
  \vert\Psi_1\rangle= \left(\matrix{1\cr 0\cr 0\cr}\right)\quad ,
  \quad \vert\Psi_2\rangle= \left(\matrix{0\cr 1\cr 0\cr}\right)\quad ,
  \quad \vert\Psi_3\rangle= \left(\matrix{0\cr 0\cr 1\cr}\right)\ .
\label{explicitnoneone}
\end{equation}
Or, alternatively,
\begin{eqnarray}
  &&\hat n(1)=\left(0,0,{\sqrt{3}\over 2},0,0,0,0,{1\over
   2}\right)\nonumber\\
  &&\hat n(2)=\left(0,0,-{\sqrt{3}\over 2},0,0,0,0,{1\over
   2}\right)\nonumber\\
  &&\hat n(3)=\left(0,0,0,0,0,0,0,-1\right)
\label{explicitonetwo}
\end{eqnarray}

We are now in a position to appreciate the advantage of resorting to
a Bloch-like parametrization. It is easier to deal with Eq.
(\ref{firstset}) than with Eq. (\ref{previouseq}). The use
of $\hat n(r)$ introduces a simple covariant, yet constrained,
formulation.  Some extra subtleties will play a relevant role in
the more complicated cases.

\section{Optimal measurements for the $N=2$ case}

Let us face the case where $N=2$ identical copies of the system
are at our disposal. Following the same reasoning as before we
start by writing Eq. (\ref{basiceq})
in terms of the basic spinor representation. This leads to
\begin{eqnarray}
  &&\sum_{r=1}^n c_r^2 \ \left(x_i(r)x_j(r)+y_i(r) y_j(r)\right)
   \left(x_k(r)x_l(r)+y_k(r) y_l(r)\right)={1\over 4}\left(
   2 \delta_{ij}\delta_{kl}+\delta_{ik}\delta_{jl}+\delta_{il}\delta_{jk}
   \right)\nonumber\\
  &&\sum_{r=1}^n c_r^2 \ \left(x_i(r)y_j(r)-x_j(r) y_i(r)\right)
   \left(x_k(r)y_l(r)-x_l(r) y_k(r)\right)={1\over 4}\left(\delta_{ik}
   \delta_{jl}-\delta_{il}\delta_{jk}\right)   \nonumber\\
  &&\sum_{r=1}^n c_r^2 \ \left(x_i(r)x_j(r)+y_i(r) y_j(r)\right)
   \left(x_k(r)y_l(r)-x_l(r) y_k(r)\right)=0
\label{secondprevious}
\end{eqnarray}
The system is now quadratic in the basic structures appearing linearly 
in the $N=1$ case. Using the Bloch vector representation,
these  $(2 J+1)^2 (2J^2+2J+1)$ equations can be recast into
\begin{eqnarray}
  &&\sum^n_{r=1} \ c^2_r \ = (2J+1)(J+1)\equiv B \nonumber\\
  &&\sum^n_{r=1} \ c^2_r \  n_a(r)=0 \nonumber\\
  &&\sum^n_{r=1} \ c^2_r \  n_a(r) n_b(r) =B\ {1\over 4J(J+1)}\ 
  \delta_{ab} 
\label{secondset}
\end{eqnarray}
A general pattern is emerging. Higher $N$ optimal measurements demand
a finer grained resolution of the identity. The Bloch vectors
are required to satisfy isotropy conditions in $SU(2 J+1)$ group space.
The determination of the factor ${1 \over 4J(J+1)}$ has been done using
that $\hat n$ is a normalized vector 
and Eq. (\ref{sumacr}). It is easy
to verify that the set of equations (\ref{secondset}) provides a solution
for Eq. (\ref{blocheq}).

From the above basic set of equations it is easy to get
\begin{eqnarray}
  &&\sum^n_{r\not= s} \ c^2_r \ = B -c_s^2 \nonumber\\
  &&\sum^n_{r\not= s} \ c^2_r \  \hat n(r)\cdot \hat n(s)=-c_s^2 \nonumber\\
  &&\sum^n_{r\not= s} \ c^2_r \  (\hat n(r)\cdot \hat n(s))^2 =B\ {1\over 4J(J+1)}- c_s^2\  
\label{secondsetbis}
\end{eqnarray}
Then we may argue that
\begin{equation}
  \sum_{r\not= s} \ c_r^2\ \left(b + \hat n(r)\cdot \hat n(s)\right)^2 \geq 0\ .
\label{trick}
\end{equation}
which is extremized by $b={c_s^2\over B-c_s^2}$ leading to
\begin{equation}
  n\geq (2J+1)^2 \qquad,\qquad c_s^2\leq {J+1\over 2J+1} \ \ \forall s
\label{boundsntwo}
\end{equation}
For $J={1\over 2}$ this bound agrees with the known solution of the
tetrahedron (see Introduction and Ref. \cite{LPT}), and generalizes it in 
the following sense. The solution $n=(2 J+1)^2$ also forces all
scalar products to be $\hat n(r)\cdot\hat n(s)=-{1\over 4 J(J+1)}$.
This corresponds to a hypertetrahedron in $(2J+1)^2-1$ dimensions,
exactly those of the adjoint representation of $SU(2 J+1)$. Let us
just write the explicit solution for $J=1$
\begin{eqnarray}
  &&\hat n(1)=\left({1\over 2},{\sqrt{3}\over 2},0,0,0,0,0,0\right)\nonumber\\
  &&\hat n(2)=\left({1\over 2},-{\sqrt{3}\over 4},{3\over 4},0,0,0,0,0\right)\nonumber\\
  &&\hat n(3)=\left({1\over 2},-{\sqrt{3}\over 4},-{3\over 4},0,0,0,0,0\right)\nonumber\\
  &&\hat n(4)=\left(-{1\over 4},0,0,0,{\sqrt{6}\over 4},{\sqrt{3}\over
   4},-{\sqrt{6}\over 4},0\right)\nonumber\\
  &&\hat n(5)=\left(-{1\over 4},0,0,{3\sqrt{2}\over 8},-{\sqrt{6}\over 8}
   ,{\sqrt{3}\over
   4},{\sqrt{6}\over 8},-{3\sqrt{2}\over 8}\right)\nonumber\\
  &&\hat n(6)=\left(-{1\over 4},0,0,-{3\sqrt{2}\over 8},-{\sqrt{6}\over 8}
   ,{\sqrt{3}\over
   4},{\sqrt{6}\over 8},{3\sqrt{2}\over 8}\right)\nonumber\\
  &&\hat n(7)=\left(-{1\over 4},0,0,0,{\sqrt{6}\over 4},-{\sqrt{3}\over
   4},{\sqrt{6}\over 4},0\right)\nonumber\\
  &&\hat n(8)=\left(-{1\over 4},0,0,-{3\sqrt{2}\over 8},-{\sqrt{6}\over 8}
   ,-{\sqrt{3}\over
   4},-{\sqrt{6}\over 8},-{3\sqrt{2}\over 8}\right)\nonumber\\
  &&\hat n(9)=\left(-{1\over 4},0,0,{3\sqrt{2}\over 8},-{\sqrt{6}\over 8}
   ,-{\sqrt{3}\over
   4},-{\sqrt{6}\over 8},{3\sqrt{2}\over 8}\right)
\label{explicittwotwo}
\end{eqnarray} 
There is still the need to perform the non-obvious step of finding
 out whether this solution
does correspond to a set of spin 1 states. For completeness we  give
this final form of the solution, that is the explicit states
$\vert\Psi_1\rangle$ through $\vert\Psi_9\rangle$ which form
the POVM,
\begin{equation}
  \vert\Psi_1\rangle= \left(\matrix{1\cr 0\cr 0\cr}\right)\quad
  \vert\Psi_2\rangle= \left(\matrix{{1\over 2}\cr 
         {\sqrt{3}\over 2}\cr 0\cr}\right)\quad
  \vert\Psi_3\rangle= \left(\matrix{{1\over 2}\cr 
         -{\sqrt{3}\over 2}\cr 0\cr}\right)
\nonumber
\label{explicitketsone}
\end{equation}
\begin{equation}
  \vert\Psi_4\rangle= \left(\matrix{{1\over 2}\cr {\rm i}
         {1\over 2}\cr {1\over \sqrt{2}}\cr}\right)\quad
  \vert\Psi_5\rangle= \left(\matrix{{1\over 2}\cr 
          {\rm i}{1\over 2}\cr -{1\over 2 \sqrt{2}}
           +{\rm i}{\sqrt{3}\over 2\sqrt{2}}\cr}\right)\quad
  \vert\Psi_6\rangle= \left(\matrix{{1\over 2}\cr 
           {\rm i}{1\over 2}\cr -{1\over 2 \sqrt{2}}-
          {\rm i}{\sqrt{3}\over 2\sqrt{2}}\cr}\right)
\nonumber
\label{explicitketstwo}
\end{equation}

\begin{equation}
  \vert\Psi_7\rangle= \left(\matrix{{1\over 2}\cr -
      {\rm i}{1\over 2}\cr {1\over \sqrt{2}}\cr}\right)\quad
  \vert\Psi_8\rangle= \left(\matrix{{1\over 2}\cr -
      {\rm i}{1\over 2}\cr -{1\over 2 \sqrt{2}}+{\rm i}
       {\sqrt{3}\over 2\sqrt{2}}\cr}\right)\quad
  \vert\Psi_9\rangle= \left(\matrix{{1\over 2}\cr -
      {\rm i}{1\over 2}\cr -{1\over 2 \sqrt{2}}-{\rm i}
      {\sqrt{3}\over 2\sqrt{2}}\cr}\right)
\label{explicitketsthree}
\end{equation}
Note that all the spinors have scalar products 
with modulus equal to ${1\over 2}$.

\section{Optimal measurements for the  $N=3$ case}

The systematics of our approach are already set. It is, though, in
the case of three copies where a major difference between spin ${1\over 2}$
and higher spin systems appears. Following an analogous reasoning to
the one in the previous sections we get
\begin{eqnarray}
  &&\sum^n_{r=1} \ c^2_r \ = {(2J+3)!\over 3! (2J)!}\equiv C \nonumber\\
  &&\sum^n_{r=1} \ c^2_r \  n_a(r)=0 \nonumber\\
  &&\sum^n_{r=1} \ c^2_r \  n_a(r) n_b(r) =C\ {1\over 4J(J+1)}\ 
  \delta_{ab}\nonumber\\
  &&\sum^n_{r=1} \ c^2_r \  n_a(r) n_b(r) n_c(r) =C\ {1\over 4J(J+1)
      (2 J+3)}\ \left({2J+1\over J}\right)^{1\over 2} d_{abc} 
\label{thirdset}
\end{eqnarray}
We have used Eqs. (\ref{sumacr}), (\ref{dprop}) and (\ref{dnnn}) for 
determining the factor 
${1\over 4J(J+1)(2 J+3)}\left({2J+1\over J}\right)^{1\over 2}$.
Again it is easy to prove that Eq. (\ref{thirdset}) verify 
Eq. (\ref{blocheq}).

For the first time the r.h.s. of one of the equations displays
a tensor structure based on the $d$-symbol. Such a term would
vanish for $J={1\over 2}$ due to the simpler structure of $SU(2)$, but
is expected for higher spins (note that the conditions
(\ref{dnnn}) are zero for spin $1\over 2$).

A bound on the number of projectors appearing in a optimal POVM
can be obtained following the by now standard procedure of
investigating manifestly positive combinations. In this case,
starting from
\begin{equation}
  \sum_{r\not= s} \left({1\over 2J}+\hat n(r)\cdot \hat n(s)\right)
  \left( b+\hat n(r)\cdot \hat n(s)\right)^2 \geq 0\ ,
\label{startnthree}
\end{equation}
one gets
\begin{equation}
  n\geq (J+1) (2 J+1)^2
\label{nthreebound}
\end{equation}
and $c_s^2\leq {(2J+3)\over 3(2J+1)}$.  That is, 
$n\geq 6$ for spin ${1\over 2}$ (which agrees with the known
result in Ref. \cite{LPT}), 
$n\geq 18$ for spin 1, $n\geq 40$ for spin ${3\over 2}$, etc.
Saturating this bound is impossible for
certain cases as implied by the following 
simple argument. If the bound were to be saturated, then Eq. 
(\ref{startnthree}) becomes a restricting condition for all scalar 
products. Indeed, $\hat n(r)\cdot \hat n(s)$ is either 
$-{1\over 2J}$ or else ${2J-1\over 2 J (2J+3)}$ for any pair
$r\not= s$. If we fix any $s$ and 
assume that the minimal solution carries $p$ scalar products of the
first type and q of the second, it follows that Eq. (\ref{thirdset})
imposes $p={1\over 2}J(2J+1)^2$ and $q={1\over 2} J(2J+3)^2$. 
For any $J$ half integer or even this causes no problem but for 
odd integer values of the spin this leads to non-integer pairs,
which is absurd. Thus, in such a case, the bound cannot be saturated.

\section{Conclusions}

We have presented explicit solutions for minimal optimal
POVMs acting  on arbitrary spin $J$ systems for the case when 
two copies are available. For  $N=3$ we have provided a rigorous
bound. The key idea to simplify the analysis consists in using
Bloch representation for pure arbitrary spin  states. 
These vectors do not span a naive $(2J+1)^2-1$ sphere, but rather 
an intrincate subspace defined through covariant restrictions. 
The power of such covariance makes the set of equations simple 
\begin{eqnarray}
  &&\sum^n_{r=1} \ c^2_r \ = {(2J+N)!\over N! (2J)!} \nonumber\\
  &&\sum^n_{r=1} \ c^2_r \  n_a(r)=0 \nonumber\\
  &&\sum^n_{r=1} \ c^2_r \  n_a(r) n_b(r) = {(2J+N)!\over N! (2J)!}\ {1\over 4J(J+1)}\ 
  \delta_{ab}\nonumber\\
  &&\sum^n_{r=1} \ c^2_r \  n_a(r) n_b(r) n_c(r) = {(2J+N)!\over N! (2J)!}\ {1\over 4J(J+1)
      (2 J+3)}\ \left({2J+1\over J}\right)^{1\over 2} d_{abc} \nonumber\\
  &&\dots
\label{nset}
\end{eqnarray}
In order to analyzed a given case with $N$ copies of the spin $J$
particle,
it is necessary to retain 
\begin{equation}
  {(4J(J+1)+N)!\over N! (4J(J+1)!}
\label{numbereq}
\end{equation}
equations in the system, that is as many rows in Eq. (\ref{nset}) as $N+1$.

Our results confirm the expected increase of needed projectors to build
a POVM as the spin of the system increases. The instances analyzed, that
is $N=1,2,3$,  seem to point at a dependence of the type 
\begin{equation}
  n_{min} \sim J^N\ .
\label{generaldep}
\end{equation}

\vspace*{1cm}
\section{Acknowledgments}

We are grateful to R. Tarrach and G. Vidal for 
continously sharing their  insight with us.
Financial support from CICYT, contract AEN95-0590,
and from CIRIT, contract 1996GR00066 are
acknowledged. A. A. acknowledges a grant from MEC.

\end{document}